\documentclass[review]{elsarticle}

\usepackage{lineno,hyperref}
\modulolinenumbers[5]

\journal{Information Science}
\usepackage{amssymb}
\usepackage{amsmath}
\usepackage{graphicx, subfigure}
\usepackage{epstopdf}
\usepackage{url}
\usepackage{array}
\usepackage{verbatim}
\usepackage{slashbox}
\usepackage{algorithmic}
\usepackage{algorithm}
\usepackage{multirow}
\usepackage{booktabs}
\usepackage{CJK}

\newtheorem{subsec:coding}{subsec:coding}

\begin{document}
\begin{CJK}{GBK}{fs}

\begin{frontmatter}
\title{Cognitive Information Measurements: A New Perspective }

\author[firstauthoraddr]{Min Chen}
\ead{minchen2012@hust.edu.cn}

\author[firstauthoraddr]{Yixue Hao}
\ead{yixuehao@hust.edu.cn}

\author[secondauthoraddr]{Hamid Gharavi}
\ead{hamid.gharavi@nist.gov}

\author[thirdauthoraddr]{Victor C. M. Leung}
\ead{vleung@ieee.org}

\address[firstauthoraddr]{School of Computer Science and Technology, Huazhong University of Science and Technology, Wuhan, China}
\address[secondauthoraddr]{National Institute of Standards and Technology, USA.}
\address[thirdauthoraddr]{Shenzhen University, Shenzhen, China}

\begin{abstract}
From a traditional point of view, the value of information does not change during transmission. The Shannon information theory considers information transmission as a statistical phenomenon for measuring the communication channel capacity. However, in modern communication systems, information is spontaneously embedded with a cognitive link during the transmission process, which requires a new measurement that can incorporate continuously changing information value. In this paper, we introduce the concept of cognitive information value and a method of measuring such information. We first describe the characteristics of cognitive information followed by an introduction of the concept of cognitive information in measuring information popularity. The new measurement is based on the mailbox principle in the information value chain. This is achieved by encapsulating the information as a mailbox for transmission where the cognition is continuously implemented during the transmission process. Finally, we set up a cognitive communication system based on a combination of the traditional communication system and cognitive computing. Experimental results attest to the impact of incorporating cognitive value in the performance of 5G networks.
\end{abstract}



\end{frontmatter}


\section{Introduction}

The increasing number of smartphones and multimedia services (e.g., virtual reality and augmented reality) and development of Internet of Vehicles (IoV) and Internet of Things (IoT) are expected to have a higher requirements for the storage and computing ability of future communication system. The number of users accessing the network is also expected to increase sharply to allow massive IoT participation. At the same time, the end-to-end delay will also be substantially reduced to only a few milliseconds~\cite{BFederico,SChen,CXWang}. However, the popularity of IoT could result in an explosive growth of mass data acquisition, which can put a tremendous pressure on the wireless communication network to satisfy the growing bandwidth demands. Despite many advanced features, 5G networks may not be able to meet such a high demand.

Furthermore, it is known to us that transmitted data normally contains too much redundancies and useless data. For example, continuous invariant video transmitted by video surveillance equipment~\cite{DSabella}. Therefore, communication networks are supposed to achieve an intelligent information transmission that is capable of selecting only useful information for transmission throughout the network.

Fortunately, with the emergence of cognitive computing, it is required to have a different way of information measurement to define information from the perspective of cognitive information. This is in contrast to the traditional Shannon information~\cite{Shannon}, which assumes that the information size remains unchanged during the transmission process. Under dynamic conditions, a cognitive information theory can be established as a combination of cognitive computing and information transmission.  Since the cognitive information regards the information value as one of the main indexes for evaluating the communication system, the core problem is how to measure the information value.

Bear in mind that when the information is generated,  it will evolve continuously and the value density (i.e., information value of unit bit) will become higher and higher with the cognition of the information. Eventually it will tend to converge. Meanwhile, upon receiving the information, users will further improve their understanding of the information according to their own knowledge.

Therefore, in this paper, we firstly provide the characteristics of cognitive information based on information popularity that represents an evolutionary process of cognitive information. Then, to realize information cognition, we put forward to the mailbox theory and its mathematical description. Finally, we propose a cognitive communication system as a combination of the traditional communication system and information cognition.

In summary, the main contributions of this work are as follows.
\begin{itemize}
\item  Introduce the concept of cognitive information value for intelligent transmission. It includes the characteristics of cognitive information involving dynamic, polarity, evolution, convergence, and multi-view as well as a supporting analysis for information popularity.
\item Develop a method of encapsulating the information as a mailbox based on the principal of mail box terminology, which recognizes the information value continuously during transmissions.
\item Design a new communication system based on cognitive computing and traditional communication system. Under these conditions, only the most valuable data is transmitted in order to reduce delay, and energy consumptions.
\end{itemize}

The remainder of this paper is organized as follows. In Section ~\ref{sec:arch}, the evolution of cognitive information is introduced. Information measurement based on cognitive computing is given in Section ~\ref{sec:tech}. In Section~\ref{sec:design}, the mailbox theory for cognitive information is presented. In Section~\ref{sec:5}, the cognitive information communication system is introduced. Finally, the conclusions are given in Section~\ref{sec.conclusion}.

\section{Evolution of Cognitive Information}
\label{sec:arch}

In this section, we provide the motivation for using cognitive information and the importance of information cognition from the perspectives of network and data.

\subsection{Motivation}

The Internet Data Center (IDC) forecasts that the Global DataSphere will grow from 33 ZB in 2018 to 175 ZB by 2025~\cite{IDC}. Thus, the processing for such a massive amount of data is quite challenging and would require sufficient storage and computing power. More importantly, it is required to develop an intelligent processing method capable of recognizing the meaningful information in real-time, which can be achieved best via information cognition.  For instance, in AlphaGo, the game of Go uses the technology of deep learning, reinforcement learning, and Monte Carlo tree search, which would require training 30,000,000 sets of human data and 48 TPU distributed systems to defeat the Lee Sedol~\cite{wang2016does}. Thus, training AlphaGo requires a huge amount of data.

However, the game of Go has to follow an accurate game rule (i.e., cognitive information). On this basis, AlphaGo Zero does not use prior knowledge nor raw data. Its learning process is based on random games, such as using self-game reinforcement learning, rather than the human data. It takes black and white pieces on a chessboard as an input and applies neural networks to simultaneously represent policy and value, rather than using them separately, as in the case of human participants. To achieve a better performance than AlphaGo, AlphaGo Zero uses 64 GPU workstations and 19 CPU parameter servers for training, and 4 TPU to perform matching execution. Thus, the rules of Go are cognitive information, which plays an important role in establishing strong artificial intelligence.

Specifically, from the perspective of cognitive information, AlphaGo training requires a huge amount of data, but the quantity of cognitive information is very low.  Thus, it is essential to extract the value of the information so that the entire data can be simplified to a small amount. Therefore, the main objective of measuring the information capacity from the perspective of cognitive computing, is to reduce the amount of information as much as possible without compromising the basic rules of information theory. In this way, a new cognition of measuring the information value (e.g., a set of rules) is needed to reduce the amount of data.

\subsection{Aspect of Cognitive Information}

The data is first generated on a user terminal and then transmitted to the remote server by a communication network, where it is coded, analyzed, and processed. The remote server feeds the processed results back to the user in the same manner via different links. In this work, we use a data transmission network and a processing terminal to realize the information cognition. It specifically consists of two parts: network cognitive optimization and data cognition.

\emph {Network cognitive optimization}: The network cognitive optimization is mainly concerned with resource distribution, scheduling, and management in order to ensure the quality of service (QoS). The main objective is to create a suitable environment for cognitive information transmission to achieve a high reliability and low delay. It specifically aims at maximizing the mutual benefits be-tween users and providers. A suitable optimization model can be used to enhance the systems' throughput and spectral efficiency, as well as simultaneously minimizing the delay and energy consumption. Possible decision configurations include transmission power distribution, resource distributing and scheduling, routing decision, spectrum resource distribution, and other processes.

Generally, network optimization consists of the following stages: i) performing data acquisition and analysis of the existing network on the premise of a full understanding of the networks' running state; ii) discovering factors that can influence network quality, which can then be optimized using various methods such as machine learning, game theory, and other methods; iii) making a network achieve its best running state by optimizing the resources.

\emph {Data cognition}: Facing various types of information, humans first filter the received data to capture valuable information, and then evaluate it in the cerebral cortex. Similarly, for communication systems in the face of dealing with mass structural and non-structural data, a system need to acquire valuable information to achieve the same goal. In traditional data processing methods~\cite{buczak2016survey}, it is required to extract the data features manually. However, in an era of big data with booming machine learning, deep learning technologies can be exploited to extract important data automatically and  achieve  better prediction~\cite{lecun2015deep}.

In deep learning, modeling and optimization would require recognition and understanding of mass data, so the volume of data is a key to improving the accuracy of the model forecasting.  Therefore, it is important to realize cognitive filtration and carry out certain screenings and discriminations before introducing the data into the construction model. This is mainly to prevent dirty data from impacting the model performance.

\section{Information Value Measurement based on Cognitive Computing}
\label{sec:tech}

In this section, we first describe the characteristics of cognitive information before presenting the measurement of information value from the perspective of information popularity.

\subsection{Characteristics of Cognitive Information}

\begin{figure}[!ht]
\centering
\includegraphics[width=\columnwidth]{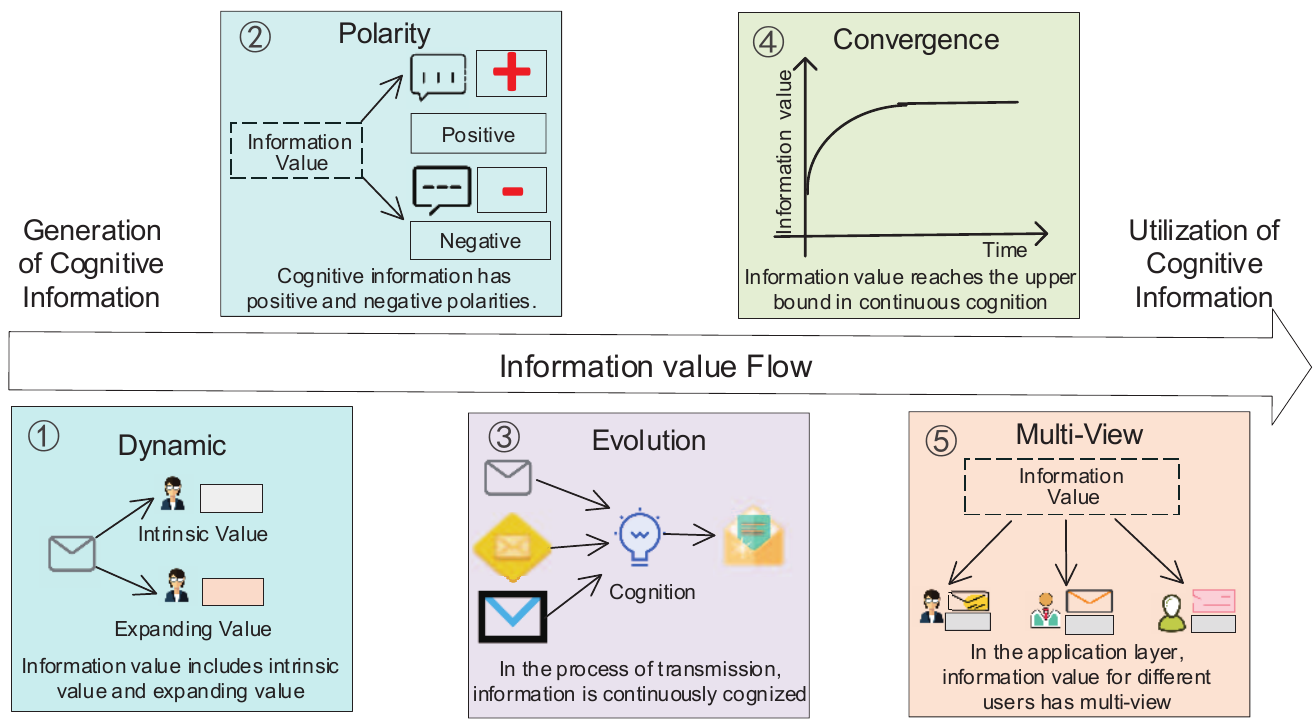}
\caption{The characteristics of cognitive information.}
\label{fig01}
\end{figure}

According to the basic concept of information, each phase in the life cycle of information possesses a certain cognitive potential, which can be utilized to assess its information value. A value, which is contained in each piece of information, can be extracted to represent a quantified measure of cognitive information. It should be noted that cognitive information expands beyond communication systems and can be used as in combination with cognitive computing to solve many research problems, such as information acquisition, transmission, analysis, and application.

To be specific, information cognition is different from conventional information measurement. We describe information cognition as follows: when information is generated, it will be recognized by different people during transmission. Therefore, after being recognized, information eventually achieves the highest value density as it evolves. The characteristics of cognitive information are shown in Fig.~\ref{fig01} and described as follows:

\begin{itemize}
\item \textbf{Dynamic:} As shown in Fig.~\ref{fig01}, information value is comprised of an intrinsic value and an expanding value. The intrinsic value denotes the intrinsic natural attribute at the beginning of the information generation. The expanding value denotes a gradually formed social attribute under the influence of external factors during the information transmission process. After generation, the information is endowed with the intrinsic value of initial information, which is constantly transmitted to different users based on their own interpretations.  During this process, the information is commented and subsequently an expanded information value is generated. As for a good information block, its potential is rated differently by each user where its value is fully excavated, analyzed, and then utilized.  It is noted that each user may have a different association and different understanding of the same information block at different times. Thus, the cognitive information value is dynamically changing during the transmission process.

\item \textbf{Polarity:} In traditional information theory, the measurement of information (i.e., Shannon entropy) is non-negative, but it possesses positive and negative polarities from the expanding value of cognitive information. During the transmission process and interaction with users, the associated information can be generated by using cognitive computing to reflect the expanding information value. It should be noted that certain information can have a positive impact on some users, but if wasteful, can have a negative impact on information transmission. For instance, for a primary school student learning that information relevant to scientific knowledge will play an important role in his/her growth and development will manifest the positive polarity of information. Conversely, if a primary school student focuses on illogical and meaningless things, it may have a negative impact on his/her cognition, manifesting the negative polarity of information. Thus, the cognition of information polarity is vitally important.

\item \textbf{Evolution:} Once the information is produced, it will be continuously recognized during the transmission process. When the cognitive capability reaches a certain level (by analogy) the information can be transformed between different dimensions. After transformation it can be applied to the data level of other dimensions to generate new information and viewpoint. The cognitive system simulates human thinking during the training process, constantly enhancing intelligence through continuous learning, which gradually approaches the cognitive capability of humans. During the process of information transmission, the valuable information is expanded and then attenuated based on the definite principles of the cognitive system, to enable the information to better meet the individual needs of users in multi-dimensional space. Thus, information possesses the evolution characteristic.

\item \textbf{Convergence:} When the information is recognized, to some extent its value density tends to be stable from the value of the information, which illustrates that information has a convergence nature. For example, the information about the description of the motion of an object can be summarized as Newton's three laws of motion. To be specific, according to the Shannon information theory, the amount of information transmitted through a digital communication system in a time unit is restricted by the channel capacity. However, in practice, demands for high volume continuous information are always at odds with the physical channel capacity. Thus, it is required to continuously improve the communication system capability and enhance data transmission volume. Furthermore, by means of recognizing the information continually it will achieve the highest value density, which reduces the transmission of information. Accordingly, to effectively reduce a burden on the communication system, the redundant data should be removed based on its information value. Namely, the removal of redundant information and conciseness of valuable information are not infinite and abide by the convergence principle on the premise of meeting the minimum reduction rule.

\item \textbf{Multi-view:} From the user's perspective, the cognitive information will reach the largest value density when it converges. Upon receiving the information, the value of the same information can impact each user differently due to users' different cognitive capabilities and demands. For instance, a high-level research paper can inspire researchers in related fields greatly and accordingly open a new research field; on the other hand, for a person who has poor scientific knowledge, such information has almost no value. This example clearly represents the information impact on different users. Since information has different convergence and value for different users, information has a multi-view nature.
\end{itemize}

\subsection{Information Popularity}

In the last section, we give the characteristics of cognitive information. In this section, for multimedia information (e.g., audio-video data), we can use information popularity to give its measurement.
To describe information popularity, we firstly introduce a definition of information. We assume that information $i$ can be described by its size $s_{i}$ (given in bits) and the required computing resource (given in CPU cycles). To analyze the information value, we begin with defining the information life cycle. Assuming that $t_{i}(t)$ is the time instance in which information $i$ is generated, then the information life cycle at the moment $t$  is given by:
\begin{equation}
\eta_{i}(t)=t-t_{i}(t).
\end{equation}

We assume that the value of information $i$ at time $t$ is determined by function $h_{i}^{t}$. We further assume that information popularity is relevant to: information size, information life cycle, and information value. Thus, the popularity of information $i$  at time $t$  is defined as:
\begin{equation}
p_{i}^{t}=f(s_{i},\eta_{i}^{t},h_{i}^{t}).
\end{equation}
where $f(\cdot)$ denotes the function relevant to $s_{i}, \eta_{i}^{t}, h_{i}^{t}$.

\section{Mailbox Theory for Cognitive Information}
\label{sec:design}

In this section, we introduce the concept of the mailbox theory for cognitive information. This includes an overview of mailbox theory and its main characteristics

\begin{figure}[!ht]
\centering
\includegraphics[width=\columnwidth]{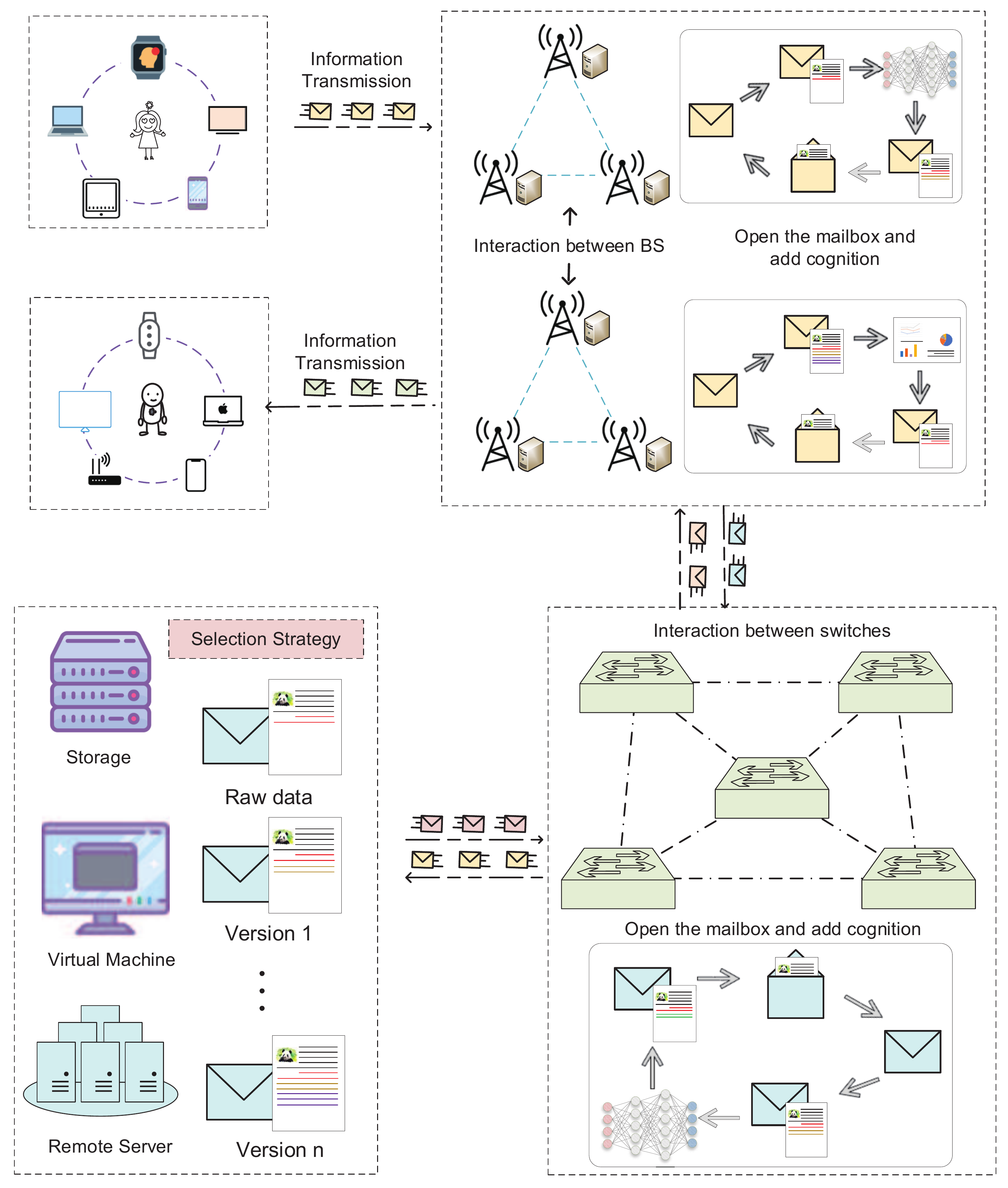}
\caption{The mailbox theory for cognitive information.}
\label{fig02}
\end{figure}

\subsection{Overview of Mailbox Theory}

In the traditional information transmission process, only the redundant bytes will be added (based on the underlying data) in order to ensure transmission reliability. However, such a scenario is not user-centered, and it neglects the information value. Mailbox theory incorporates cognition of the information value in the transmission process. It counts on the realization of the  user-centered information transmission by means of information cognition and filtering out valueless information where only higher value density information can be transmitted.

For example, we apply the mailbox to the multimedia communication for the emotion detection application. In the emotion detection application ~\cite{zhang2017learning,soleymani2015multimodal}, mobile users realize emotional detection by transmitting audio-video data to the cloud. Our goal is to improve the accuracy of the emotion recognition model in the cloud by collecting the emotion recognition data of the terminal. According to our proposed mailbox terminology, we only transmit data that can be employed to improve the accuracy of the model, hence reducing data transmission.

Specifically, we consider that information transmission is a dynamic process with entropy production~\cite{MOizumi}. Furthermore, to understand information cognition, we consider data generated by a user as an encapsulated mailbox, rather than static information. Also, we assume that users have different understandings of the information through reading and analysis. For instance, a user can perform a labeling operation on each piece of information by adding its own cognition and judgment.  As an example, we consider emotional data transmission where each user can evaluate the contribution of the data and its possible impact on the accuracy of its emotion model. Based on  recognition of the information, it will then decide whether or not to transmit the data. Data such as text, voice, or image, represents a mailbox where each mailbox has a different dimension and the information generated by interpretation is different for each user.

Next, we describe the detailed process of mailbox theory based on the example shown in Fig.~\ref{fig02}. We assume that a batch of data at the source is encapsulated into a mailbox for transmission. Normally, in a communication network data packets may move along different nodes through their paths before reaching their destinations. A node with computing resources can open its mailbox and read its content to learn about the adopted policy and the algorithm in its path, and the information value.  To guarantee high transmission efficiency generally, it is required to use an algorithm with lower complexity. Since high transmission efficiency and lower complexity are two contradictory optimization objectives, it is required to design algorithms to balance between the two.

After completing the information value measurement (e.g., the contribution of data to the algorithm in the emotion recognition model) at each node, the comments are dynamically added and then forwarded to the next node. After each node, including the raw data and the comments added by previous nodes are evaluated and then forwarded to the next node (if it decides not to drop it). Finally, at the destination node, the raw information and all comments are added to the data packet throughout its communicant link and can be analyzed.

\subsection{Introduction of Mailbox Theory}

We encapsulate the information into a mailbox, which can be then transmitted through the communication system. In the process of transmitting mailbox, due to the added comments in the packets, the mailbox size is changing at each node. Therefore, the end user (or server) is expected to receive data packets of different sizes. Moreover, due to dynamically adding cognitive comments to the raw information at each node the information quantity of each data packet is continually increasing. It is worth noting that during the mailbox information transmission the data packet can be unpacked directly at different network nodes. However, in order to protect information transmission security, the comments on data can be added to the raw data only if a node has permission to modify the packet. This is mainly to ensure the integrity and safety of raw data. Thus, every node can gain the raw information transmitted from the source. Furthermore, when a data packet reaches the terminal, various characteristics of information (e.g., encapsulation format of data packets, coding and de-coding methods, etc.) should be consistent with those of the source-transmitting end. A series of processes for packet unpacking and adding comments are experienced during the transmission process, but the basic characteristics of the data packet must be consistent.

To describe a mailbox, we assume that a mailbox is defined by two important indexes: information value, and information size. These two indexes can be changed during the information transmission process. The information value of an encapsulated mailbox $i$ at time instance $t$, is defined by function $h_{i}^{t}$. Assume that $n$ users label the information during the transmission process (i.e., from time $t$  to $t+1$). Also, assume that the change in the information value due to the labeling of a user $j$ is $Y_{j}$. Thus, the information value at $t+1$ is given by:
\begin{equation}
h_{i}^{t+1}=h_{i}^{t}+\sum_{j=1}^{n}Y_{j}
\end{equation}

Thus, the average information value of information $i$ from its generation to moment $t$ is defined by:
\begin{equation}
\overline{h}_{i}=\frac{1}{T}\mathbb{E}\left[\int_{0}^{T}h_{i}^{t}dt\right]
\end{equation}
Furthermore, the information value can be expressed by:
\begin{equation}
\overline{h}_{i}=\lim_{T\rightarrow\infty}\frac{1}{T}\mathbb{E}\left[\int_{0}^{T}h_{i}^{t}dt\right]
\end{equation}
Similarly, we assume that the information size of information $i$ at time instant $t$ is $s_{t}$, and that $n$ users label the information during transmission , such that the size of each label is $x_{j}$. Thus, the information size at $t+1$  is given by:
\begin{equation}
s_{i}^{t+1}=s_{i}^{t}+\sum_{j=1}^{n}x_{j}
\end{equation}
Then, the ultimate information size is expressed by:
\begin{equation}
\overline{s}_{i}=\lim_{T\rightarrow\infty}\frac{1}{T}\mathbb{E}\left[\int_{0}^{T}s_{i}^{t}dt\right]
\end{equation}

\section{Cognitive Information based Communication System}
\label{sec:5}

Considering the user requests for various services, traditional communication technology fails to meet such high communication demands. Thus, in this section, in order to optimize existing communication systems, we propose a new communication system based on cognitive information by exploiting the mailbox theory. The proposed cognitive information based communication system is aimed at multimedia communication, which takes into account that the existing multimedia data (e.g., audio and video data) is the main traffic in the existing network.

\subsection{Proposed communication system}

\begin{figure}[!ht]
\centering
\includegraphics[width=\columnwidth]{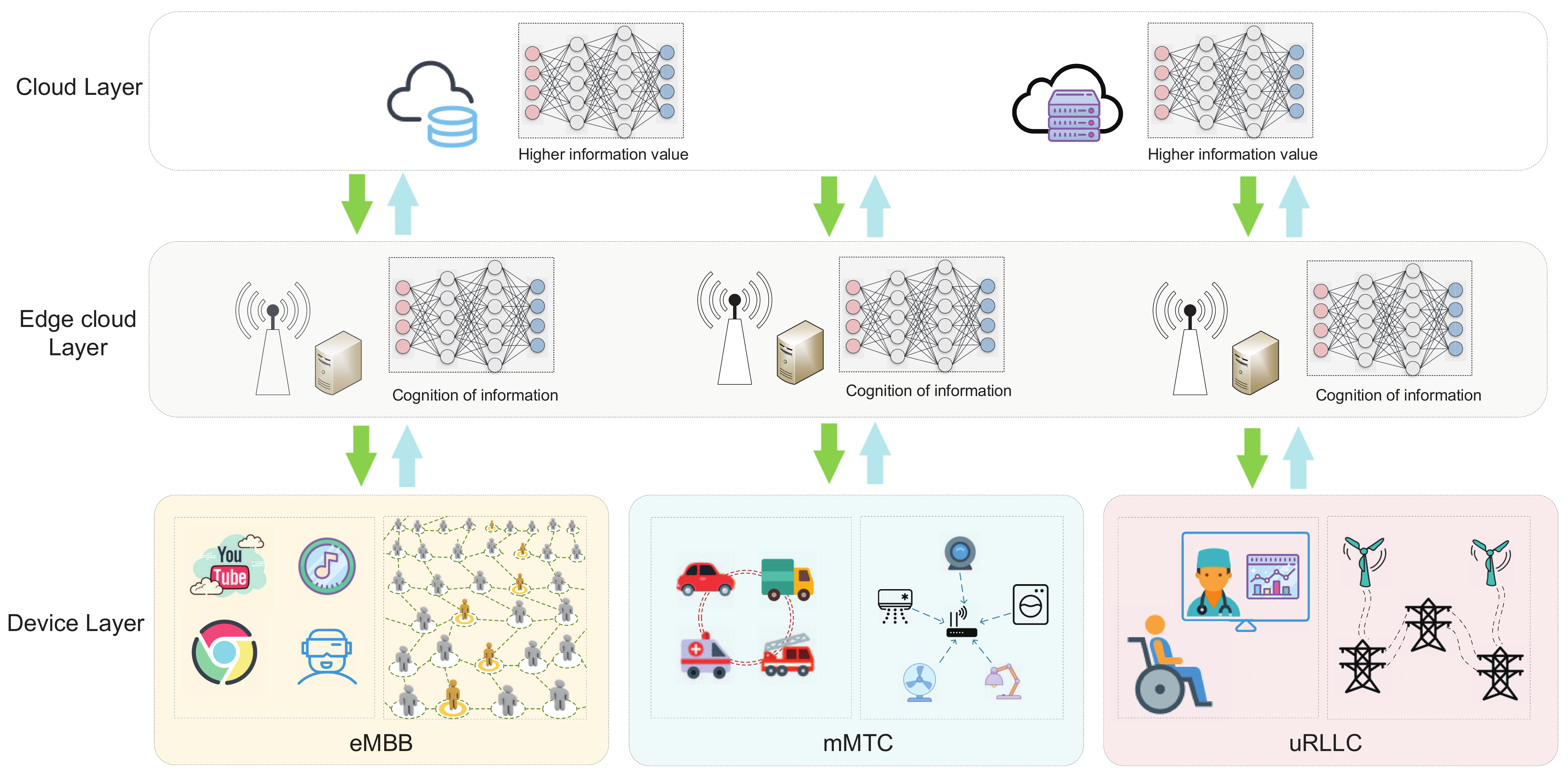}
\caption{The architecture of the proposed communication system.}
\label{fig03}
\end{figure}

In this paper, we construct a novel communication system for transmission of emotion data that is based on mailbox theory. In the proposed communication system, the main objective of mailbox terminology is to reduce the data transmission by classifying the information by which only valuable information can be selected for transmission. To achieve it, we use cognitive information based on machine learning to obtain the information value. The architecture of the proposed communication system, which includes device layer, edge layer, and cloud layer, is shown in Fig.~\ref{fig03}. Taking emotional recognition as an example, we encapsulate face expression data and voice data as mailbox. The data includes labeled and unlabeled data. Compared with the traditional method of offloading all data to the cloud, emotional recognition is directly carried out in the cloud. At first we recognize the information and after selecting and identifying the unlabeled data we offload optimal valuable data to the cloud. Thus, by recognizing the information, the intelligence and reliability of the proposed system can be realized.

In the following we will show how emotion information is recognized. Bear in mind that in traditional communication systems, all data (i.e., information) needs to be offloaded to the cloud~\cite{chen2017emotion,xu2018survey}. In our proposed communication system, only valuable data (after being identified by using machine learning) will be offloaded to the cloud~\cite{chen2018label}. It should be noted that in the emotion detection field, the accuracy of the trained model depends on the data samples quality. In the classification problem, we can easily determine the classification confidence of a data sample by using the machine learning model. Under these conditions, we can obtain the information entropy of data by using the confidence value.

\begin{figure}[!ht]
\centering
\includegraphics[width=\columnwidth]{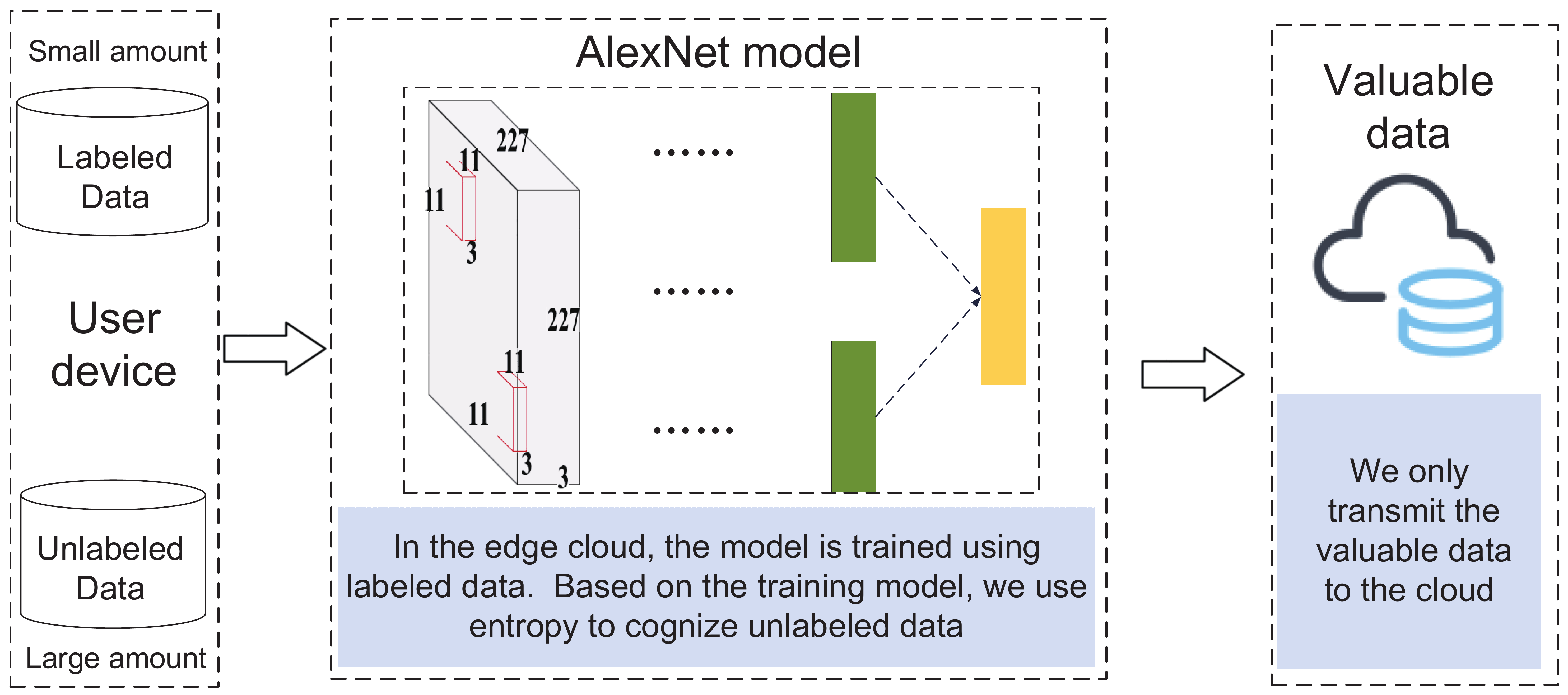}
\caption{The information value measurement during AlexNet training.}
\label{fig04}
\end{figure}

To be specific, the model framework and the training process are shown in Fig.~\ref{fig04}. In order to recognize emotions, it is required that end users transmit large amounts of data to the cloud. For the data, only a small part of data is labeled, while most of the data does not have any label (i.e., unlabeled data) and contain much noise. Therefore, we first utilize the AlexNet~\cite{krizhevsky2012imagenet} architecture which includes 5 convolution layers, 3 pooling layers and 2 full connection layers for emotion detection based on labeled data. Second, unlabeled data is used to further improve the accuracy of the emotion detection model. Assuming that the unlabeled data $x_{i}^{u}$  is predicted through AlexNet with a probability of $p_{x_{i}^{u}}=\{p_{x_{i}^{u}}^{1}, p_{x_{i}^{u}}^{2}, \cdots, p_{x_{i}^{u}}^{c}\}$, where $p_{x_{i}^{u}}^{j}$ represents the probability that unlabeled data $x_{i}^{u}$ is classified into emotion $j$. Thus, we can get the following formula:
\begin{equation}\label{eq:8}
E(p_{x_{i}^{u}})=-\sum_{j=1}^{c} p_{x_{i}^{u}}^{j}\log_{2}p_{x_{i}^{u}}^{j}.
\end{equation}
where $c$ is the number of classes. From formula~\eqref{eq:8}, we can see that when $E(p_{x_{i}^{u}})$ is small, the data has less prediction uncertainty. Thus, the unlabeled data can be classified more easily and can be more valuable to model training and parameter iteration; When $E(p_{x_{i}^{u}})$ is large, the data has greater prediction uncertainty so the contribution of this data to the model is small. Therefore, the amount of data transmission can be reduced by cognizing the data.

To enhance computing ability to recognize data, we first set up the AlexNet model on the edge server and transmit the labeled data for model training in order to obtain an immature model. Then, we select partial data from the unlabeled data and transmit it to the edge server for model training where we can compute the information entropy by using~\eqref{eq:8}. Next, we set an experiment which defines information value threshold to decide whether transmit the unlabeled data to the remote cloud based on the computed information value (if the information value is lower than the threshold, data will be transmitted; otherwise, the data packet will be abandoned). Finally, we select the unlabeled data continuously, until all the user data goes through value assessment. Thus, data screening can alleviate the flow congestion in the network for every node in the network.

\subsection{Experimental setup and performance evaluation}

\begin{figure}[!ht]
\centering
\includegraphics[width=0.5\columnwidth]{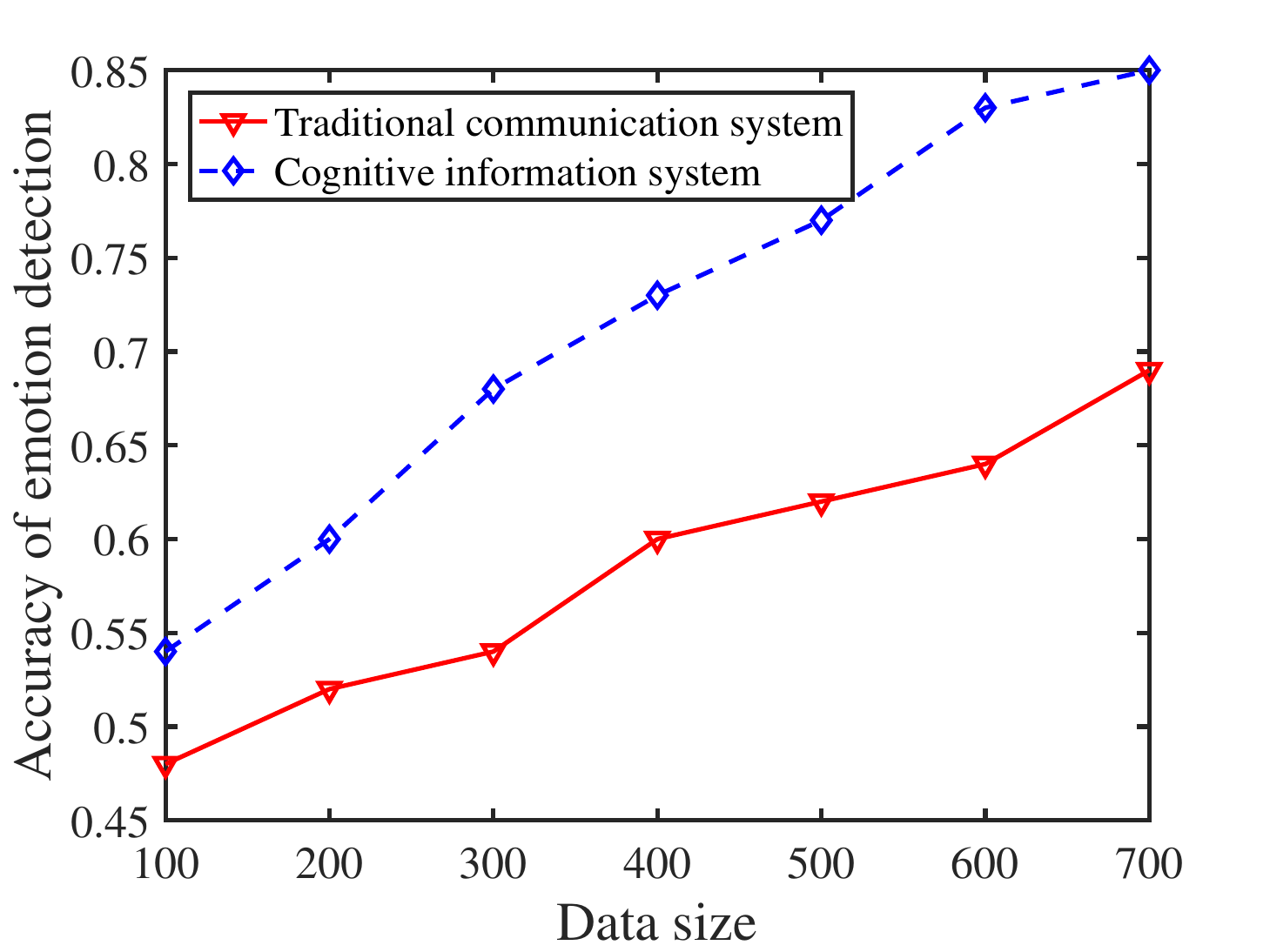}
\caption{Accuracy of emotion detection versus data size under different communication systems.}
\label{fig05}
\end{figure}

In the experiment, we encapsulate the facial expression data and speech data generated by users as mailbox, and cognize the data on the edge cloud. The value of data is measured by the contribution of emotional data to the accuracy of the model and using~\eqref{eq:8} for evaluation. On this basis, by discarding the data with low value density, we only transfer the most valuable information to the cloud emotion detection model. In our experiments, we compare the cognitive information system with the traditional information communication system, which transmits all information to the cloud without cognizing the transmitted information. To be specific, at first we compare the impact of the cognitive information system and traditional communication system on the accuracy of emotion detection model for transmitting the same amount of data.

From Fig.~\ref{fig05}, we can clearly observe that the accuracy of emotion detection in the cognitive information system is higher for a given data transmission size. The accuracy of the model refers to the proportion of the predicted data labeled with real labels to the total data. From the figure, we know that in the cognitive system, by cognizing the unlabeled data, it can transmit data with a high value density (i.e. contribution to model accuracy) to the cloud, thus improving the accuracy of emotion recognition. Generally speaking, through the cognition of information, the system proposed in this paper can transmit less data when transmitting data of the same value.

\begin{figure}[!ht]
\centering
\includegraphics[width=0.5\columnwidth]{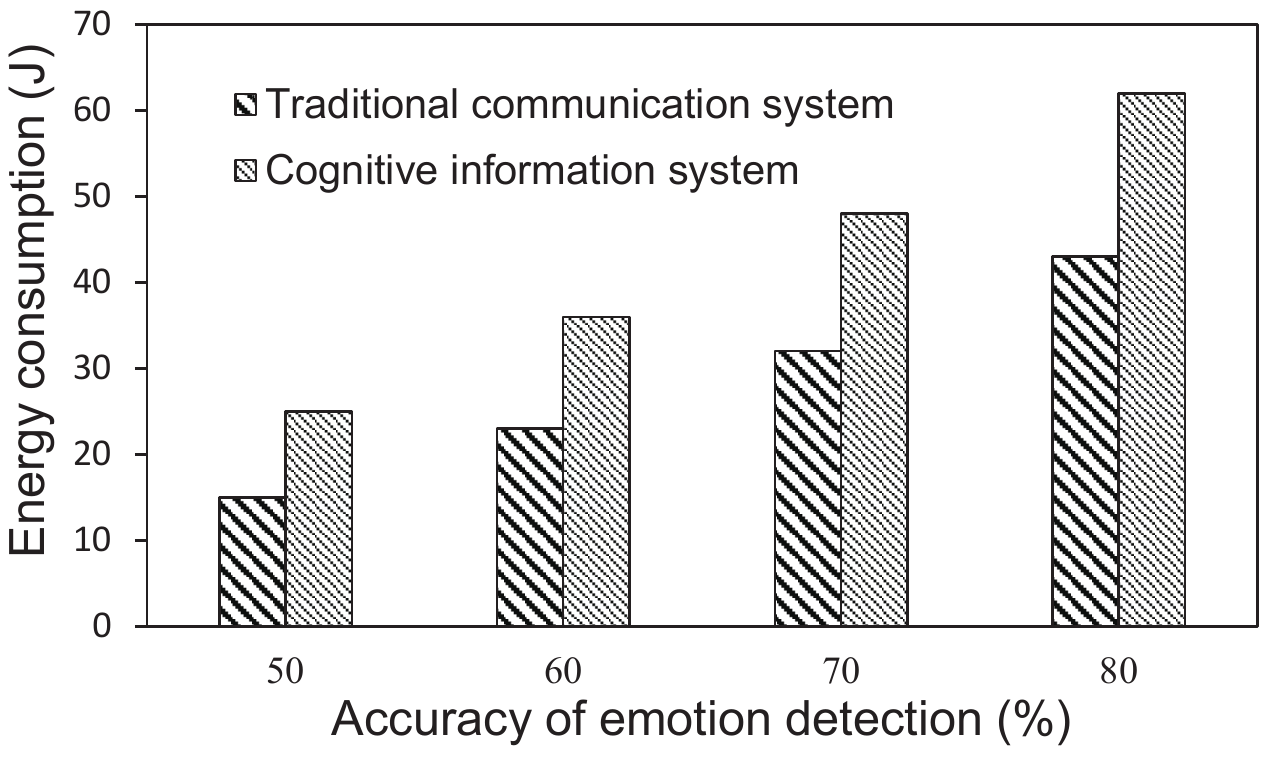}
\caption{Energy consumption versus accuracy of emotion detection under different communication systems.}
\label{fig06}
\end{figure}

Next, we evaluate the performance of the cognitive information system from the perspective of energy consumption. Compared with the traditional information system, the energy consumption of the cognitive information system not only includes transmission energy consumption, but also computing energy consumption involved in the cognitive processing of information. Fig.~\ref{fig06} shows the energy consumption of the two systems under the same accuracy of emotion detection. The experimental results show that although cognitive information system requires extra computational energy consumption, the total energy consumption of the cognitive information system is still less than the traditional communication system under the same accuracy rate for the emotional detection model. This is mainly due to the scientifically reduced transmission rate of the cognitive information system in the cloud, which can result in an overall reduction of the energy consumption.

\begin{figure}[!ht]
\centering
\includegraphics[width=0.5\columnwidth]{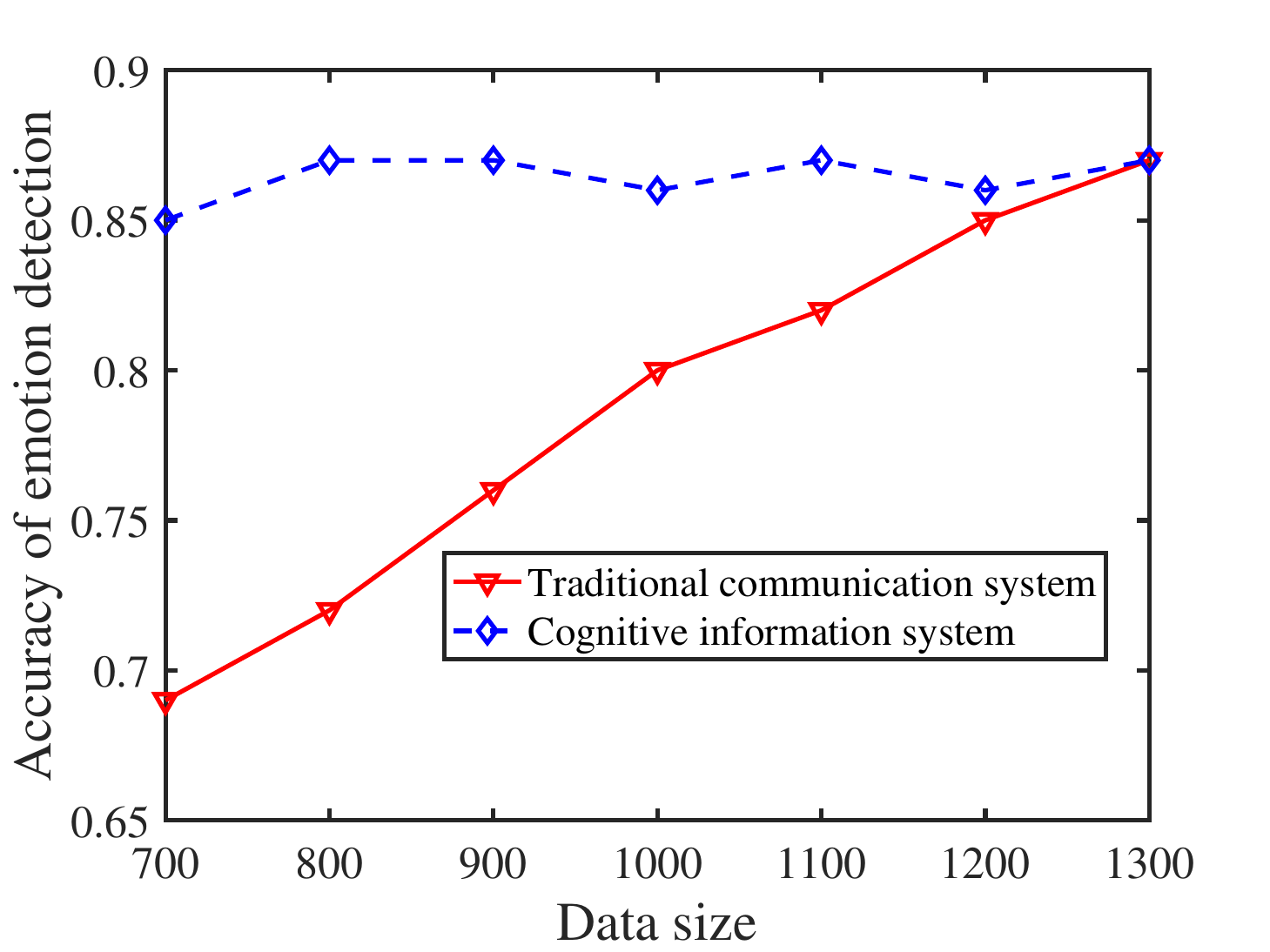}
\caption{Convergence rate of different communication systems.}
\label{fig07}
\end{figure}

Finally, we investigate on the information convergence speed. We define the convergence of data as a degree in which new unlabeled data can no longer improve the accuracy of the emotion detection model in the cloud. Fig.~\ref{fig07} shows a comparison of convergence speeds between the two systems. We can see that although the accuracy of the two systems eventually tends to become stable, the cognitive information system tends to converge faster. As soon as the value of information becomes stable, the transmission of information will be terminated.

\subsection{Discussion}

Although this scheme can realize information cognition, its deployments depends on the following technologies.
\begin{itemize}
\item Artificial intelligence chip. The computing capability of a mobile phone is constantly increasing to enable machine learning algorithms to be deployed at the mobile terminal where it can directly handle data computing and screening.

\item Edge computing. Using the storage and computing power of edge cloud, data processing can be realized on an edge server in order to guarantee low latency and alleviate the traffic congestion of a core network.

\item Network function virtualization. By using software-defined network (SDN) and network function virtualization (NFV), we can realize the coupling of control and user planes as well as the coupling of hardware and software.
\end{itemize}

\section{Conclusion}
\label{sec.conclusion}

In this paper, considering information variability, we first introduce a cognitive information theory and its characteristics involving dynamic, polarity, evolution, convergence, and multi-view. Then, we provide the mailbox theory in cognitive information. Finally, combining the traditional communication system and cognitive information theory, we propose a new communication system. The results of the experiments show that the cognitive communication system has higher performance in terms of transmission rates.


\section*{Reference}
\bibliography{reference}

%

\end{CJK}
\end{document}